\documentclass[12pt]{article}
\usepackage{epsf}
\newcommand{\refc}[1]{Ref.~\cite{#1}}

\newcommand{\bk}[2]{\left<#1|#2\right>}
\newcommand{\bko}[3]{\left<#1|#2|#3\right>}
\newcommand{\bma}[1]{\mbox{\boldmath $#1$}}
\newcommand{\be}{\begin{equation}}
\newcommand{\ee}{\end{equation}}
\newcommand{\bea}{\begin{eqnarray}}
\newcommand{\eea}{\end{eqnarray}}
\newcommand{\eqr}[1]{Eq.~(\ref{#1})}
\newcommand{\eqrs}[2]{Eqs.~(\ref{#1}) and (\ref{#2})}

\newcommand{\dih}{$\Delta I\!=\!1/2$}
\newcommand{\dit}{$\Delta I\!=\!3/2$}
\newcommand{\ktp}{$K\to2\pi$}
\newcommand{\ktpg}{$K\to2\pi\gamma$}
\newcommand{\cpvs}{\mbox{\scriptsize CPV}}
\newcommand{\cpcs}{\mbox{\scriptsize CPC}}
\newcommand{\dihs}{\mbox{\scriptsize $\Delta I\!=\!1/2$}}

\newcommand{\ttm}[1]{\times10^{-#1}}
\newcommand{\Real}[1]{\Re {\it e}\left(#1 \right)}

\begin{document}\baselineskip .7cm
\title{CP Violation and $\Delta I\!=\!1/2$ Enhancement \\ for $K\to\pi\pi$,
$K\to\pi\pi\gamma$ Weak Decays}
\author{
Michael D.\ Scadron$^{\,a}\!\!$\,,
George Rupp$^{\,b}\!\!$\,, 
and Eef van Beveren$^{\,c}\!\!$ \\[5mm]
$^{a}${\footnotesize\it Physics Department, University of Arizona, Tucson,
AZ 85721, USA} \\ {\footnotesize\tt scadron@physics.arizona.edu} \\[.3cm]
$^{b}${\footnotesize\it Centro de F\'{\i}sica das Interac\c{c}\~{o}es
Fundamentais, Instituto Superior T\'{e}cnico, P-1049-001 Lisboa, Portugal} \\
{\footnotesize{\tt george@ist.utl.pt} (corresponding author)} \\[.3cm]
$^{c}${\footnotesize\it Departamento de F\'{\i}sica, Universidade de Coimbra,
P-3004-516 Coimbra, Portugal} \\ {\footnotesize\tt eef@teor.fis.uc.pt} \\[5mm]
}

\date{\today}

\maketitle

\begin{abstract}
Data indicate that \dit\ transitions account for 4.5--4.7\% of both CP
conserving and CP violating $K\!\to\!2\pi$ decays, as well as CP conserving
radiative $K\!\to\!\pi\pi\gamma$ processes. Observed
$K\!\to\!\pi\pi\gamma/\pi\pi$ branching ratios are shown to scale near
$\alpha/\pi$ or $\alpha/2\pi$. The $K_L$-$K_S$ mixing angle $\phi$ and the
semileptonic weak-rate asymmetry $\delta$ are reviewed, and theory is shown to
be consistent with data. Also, $K\!\to\!2\pi$ \dih\ dominance is studied in the
context of the chiral constituent quark model, displaying again excellent
agreement with data. Finally, indirect and direct kaon CP violation (CPV) are
successfully described in the framework of photon-mediated quark-loop graphs.
This suggests that kaon CPV can be understood via second-order weak
transitions, radiatively corrected.
\end{abstract}

\section{Introduction}
The experimentally observed \cite{CCFT64} violation of CP in the neutral-kaon
system is, in the Standard Model (SM), parametrized through a complex phase in
the Cabibbo-Kobayashi-Maskawa (CKM) \cite{CKM63_73} quark-mixing matrix, the
origin of which is usually attributed to physics beyond the SM (see, e.g.,
Ref.~\cite{N02} for a recent analysis of possible new-physics signals).
Moreover, kaon CP violation (CPV) is well tested for \em both \em \/\ktp\ and
\ktpg\ weak transitions. Another empirically well-established phenomenon in
neutral-kaon decays is \dih\ enhancement and, likewise, \dit\ suppression.
Faced with surprising regularities in the measured branching ratios of CP
conserving (CPC) and CPV kaon decays, as well as for such processes involving a
photon, we are led to study the hypothesis that kaon CPV can be described as an
electromagnetic (e.m.) radiatively corrected second-order weak (SOW) effect.

This paper is organized as follows. In Sec.~2, we study CPV for 4.5--4.7\%
suppressed \dit\ amplitudes, in Sec.~3 the suppression factor $\alpha/\pi$
for $K\to\frac{2\pi\gamma}{2\pi}$, in Sec.~4 the CPV observed angles $\phi$
and $\delta$, in Sec.~5 a chiral quark model for \dih\ weak decays, and in
Sec.~6 both direct and indirect CPV of \ktp\ amplitudes. By bringing into the
analysis the \ktpg\ decays, CPV appears to follow from studying radiative
e.m.\ corrections to SOW transitions, together with \dih\ dominance.
\section{CPV for Suppressed \boldmath{\dit} Amplitudes}
First we review these patterns based on the recent Particle Data Group (PDG)
tables \cite{PDG02} for \ktp\ decays. Specifically, the branching ratios for
CPC and CPV \ktp\ decays are
\bea 
B\left(K_S\,\to\,\frac{\pi^+\pi^-}{\pi^0\pi^0}\right)_{\mbox{\scriptsize CPC}}
  = & \displaystyle\frac{(68.60\pm0.27)\,\%}{(31.40\pm0.27)\,\%}
& = 2.185\pm0.021 \; , 
\label{bcpc} \\[2mm]
B\left(K_L\,\to\,\frac{\pi^+\pi^-}{\pi^0\pi^0}\right)_{\mbox{\scriptsize CPV}}
 = & \displaystyle\frac{(2.084\pm0.032)\times10^{-3}}
{(9.42\pm0.19)\times10^{-4}} & = 2.212\pm0.056 \; ,
\label{bcpv}
\eea
suggesting that the mechanism driving the CPC \dih\ enhancement may also play 
a role in CPV for $K$ decays. We shall return to this point in Sec.~6.

Here we note that, since a pure \dih\ \ktp\ transition requires a branching
ratio $B(K_S\to\frac{\pi^+\pi^-}{\pi^0\pi^0})_{\mbox{\scriptsize \dih}}=2$,
the data in \eqr{bcpc} implies an approximate \dit\ amplitude contamination
of $\sqrt{2.185/2}-1=4.5$\%. Stated another way, the $K_S\to\pi^+\pi^-$ to
$K^+\to\pi^+\pi^0$ rate ratio (neglecting from now on the small experimental
errors) is, for $\tau_S=0.8935\times10^{-10}\,s$ and $\tau_{K^+}=1.2384\times
10^{-8}\,s$ \cite{PDG02},
\be
\left(\frac{\Gamma_{K_S\,\pi^+\pi^-}}{\Gamma_{K^+\,\pi^+\pi^0}}\right)_{\cpcs}
\; \approx \; \frac{\tau_{K^+}\times68.60\times10^{-2}}{\tau_{K_S}
\times21.13\times10^{-2}} \; \approx \; 449.98 \; \approx \; (0.0471)^{-2} \; ,
\label{bkskp}
\ee
implying a small \dit\ to \dih\ amplitude ratio of 4.7\%. Not only is this
small 4.7\% \dit\ suppression compatible with the above approximate CPC 4.5\%
value, but the \em radiative \em \/\dih\ to \dit\ rate ratio (divided by 2,
since the photon does not interact with a neutral $\pi^0$) \cite{PDG02}
\be
\left(\frac{\Gamma_{K_S\,\pi^+\pi^-\gamma}}{2\Gamma_{K^+\,\pi^+\pi^0\gamma}}
\right)_{\cpcs} \; \approx \; \frac{\tau_{K^+}\times1.78\times10^{-3}}
{2\tau_{K_S}\times2.75\times10^{-4}} \;\approx\; 448.56 \;=\; (0.0472)^{-2}
\label{bkskpr}
\ee
is very close to \eqr{bkskp} above. Alternatively, we could follow Okun's text
\cite{O84}, and compute the \dit\ to \dih\ amplitude ratio using Clebsch-Gordan
coefficients and the observed $\pi\pi$ phase-shift difference of
$53^\circ\pm6^\circ$ to
obtain a small 4.5--4.6\% ratio, compatible with the 4.5\%, 4.7\%, 4.71\%,
and 4.72\% values above. Also, as we shall later comment on the
theoretical version of the \dih\ rule in Sec.~5, the ratio of \eqr{hwppk}
divided by \eqr{mkspmt} again recovers the small \dit\ to \dih\ value of 4.7\%.

\section{Radiative Scale \bma{\alpha/\pi}}
Now we study the observed branching ratios of radiative relative to
non-radiative $K$ decays, and note that they are both close to the radiative
factor $\alpha/\pi=0.00232$, or $\alpha/2\pi$ if there is a $\pi^0$ in the
final state \cite{PDG02}:
\bea 
\!\!\!\!\!\!B\left(K_S\,\to\,\frac{\pi^+\pi^-\gamma}{\pi^+\pi^-}\right)_{\cpcs}
= & \displaystyle\frac{1.78\times10^{-3}}{68.60\times10^{-2}}
& = 0.00259 \;\; \mbox{vs.} \;\;\;\frac{\alpha}{\pi}=0.00232\;,
\label{brnrs} \\[2mm] 
\!\!\!\!\!\!B\left(K^+\,\to\,\frac{\pi^+\pi^0\gamma}{\pi^+\pi^0}\right)_{\cpcs}
= & \displaystyle\frac{2.75\times10^{-4}}{21.13\times10^{-2}}
& = 0.00130 \;\; \mbox{vs.} \;\; \frac{\alpha}{2\pi}=0.00116\;.
\label{brnrp}
\eea
A more detailed bremsstrahlung calculation of these branching ratios is in
good agreement with the experimental data \cite{WS86}, and thus also close to
our radiative factors of $\alpha/\pi$ and $\alpha/2\pi$.

Since the major fraction of the rate $K_L\to\pi^+\pi^-\gamma$ is CPC, it is
difficult to extract the pure CPV branching ratio
$B(K_L\to\pi^+\pi^-\gamma/\pi^+\pi^-)_{\cpvs}$. However, invoking the measured
CPV amplitude magnitudes \cite{PDG02}
\be
\!\!\!\!\!\!|\eta_{+-}| \; = \; (2.286\pm0.017)\times10^{-3} \;\;\;\;,\;\;\;\;
            |\eta_{00}| \; = \; (2.274\pm0.017)\times10^{-3} \; ,
\label{etas}
\ee
along with the radiative amplitude $|\eta_{+-\gamma}|$ of about
$2.35\times10^{-3}$, we can anticipate the above CPV branching ratio is about
\be
\hspace*{-20pt}B\left(K_L\,\to\,\frac{\pi^+\pi^-\gamma}{\pi^+\pi^-}\right)_
{\cpvs}  \! = \left|\frac{\eta_{+-\gamma}}{\eta_{+-}}\right|^2
B\left(K_S\,\to\,\frac{\pi^+\pi^-\gamma}{\pi^+\pi^-}\right)_{\cpcs} \!
\approx  2.7\times10^{-3} \; ,
\label{brnrls}
\ee
near the radiative branching ratio of \eqr{brnrs}, or the theoretical value
$\alpha/\pi=2.32\times10^{-3}$. Thus, as in Sec.~2, the branching ratios of
CPC and CPV processes are almost identical.

\section{Observed Angles \bma{\phi\,,\delta} and CPV}
Other measures of CPV are the observed angles $\phi$ and $\delta$. The
$K_L$-$K_S$ mixing angle is (assuming CPT conservation \cite{PDG02})
\be
\phi \; = \; \phi_{+-} \; = \; \phi_{00} \; = \; (43.51\pm0.06)^\circ \; .
\label{phi}
\ee
Being close to the CPC $K_L$-$K_S$ mixing angle $45^\circ$, \eqr{phi} is one
measure of CPV. Stated another way, SOW
$K^0$-$\bar{K}^0$ mixing uses the angle $\phi$ to diagonalize the mass matrix
\be
\left( \begin{array}{cc}
m^2_{K^0} & \lambda \\ \lambda & m^2_{\bar{K}^0}
\end{array} \right) 
\;\;\raisebox{-12pt}{$\stackrel{\displaystyle\longrightarrow}{\phi}$}\;\;
\left( \begin{array}{cc}
m^2_S & 0  \\ 0 & m^2_L
\end{array} \right) 
\label{diag}
\ee
via the mixing angle (near $45^\circ$) \cite{SE95,CS96}:
\be
\sin2\phi \; = \; \frac{2\lambda}{m^2_L-m^2_S} \; \approx \; 
\frac{\lambda}{\Delta m_{LS}\,m_K} \; \approx \; 1 \; .
\label{sintphi}
\ee
Using the latter equation, the angle $\phi$ can also be estimated via the
two-pion-dominated value linking the first-order weak (FOW) $K_{2\pi}$ decay
rate \cite{PDG02} $\Gamma_S\approx7.367\times10^{-12}$ MeV to the SOW
$K_L\!-\!K_S$ mass difference \cite{PDG02}
$\Delta m_{LS}=3.490\times10^{-12}$ MeV as \cite{MRR69}
\be
\phi \; = \; \arctan\!\left(\frac{2\Delta m_{LS}}{\Gamma_S}\right) \;
\approx \; 43.45^\circ \; .
\label{aphi}
\ee
We note the close proximity of \eqr{aphi} to \eqr{phi} as another ``measure''
of CPV. 

Note, too, that the radiative-induced value of $\phi$ has also been observed
\cite{PDG02}, i.e., $\phi_{\pi^+\pi^-\gamma}=(44\pm4)^\circ$, reasonably near
the non-radiative $\phi$ in \eqr{phi}, and also near the $2\pi$-dominated value
in \eqr{aphi}.

A second phase-angle measurement stems from the semileptonic weak-rate
asymmetry \cite{PDG02}
\be
\delta\;=\;\frac{\Gamma_{K_L\to\pi^-\ell^+\nu}-\Gamma_{K_L\to\pi^+\ell^-\nu}}
                {\Gamma_{K_L\to\pi^-\ell^+\nu}+\Gamma_{K_L\to\pi^+\ell^-\nu}}
\; = \; (3.27\pm0.12)\times10^{-3} \; .
\label{delta}
\ee
The model-independent relation predicting $\delta$ from $|\varepsilon|\approx
|\eta_{+-}|\approx|\eta_{00}|\approx2.3\times10^{-3}$ in \eqr{etas}, using
$\phi\approx43.5^\circ$ from Eqs.~(\ref{phi}) or (\ref{aphi}), reads
\cite{CS96}
\be
\delta \; = \; 2\,|\varepsilon|\cos\phi \; \approx \; 3.34\times10^{-3} \; ,
\label{dephi}
\ee
close to the observed $\delta$ in \eqr{delta}.

\section{Chiral Quark Model for \bma{\Delta I\!=\!1/2} Dominance}

Partially conserved axial currents (PCAC) --- consistently combined with the
chiral charge algebra $[Q+Q_5,H_w]=0$ for $H_w$ constructed from $V\!-\!A$
currents --- requires the FOW $K_S\to\pi^+\pi^-$ amplitude to satisfy
\cite{KS92}, for $f_\pi\approx93$ MeV,
\be
\left|M^{+-}_{K_S}\right|\;=\;\frac{1}{f_\pi}\,
\left|\bko{\pi^+}{H_w}{K^+}\right| \, \left(1-\frac{m^2_\pi}{m^2_K}\right) \; .
\label{mkspm}
\ee
The single-quark-line (SQL) \dih\ transition scale $\beta_w$ gives the FOW
amplitude magnitude
\be
\left|\bko{\pi^0}{H_w}{K_L}\right|\;=\;2\,\beta_w\,m^2_K\,\frac{f_K}{f_\pi}\;.
\label{hwpkl}
\ee
Also, $\beta_w$ is fixed from the SOW soft-kaon theorem of current algebra
\cite{SE95,CS96,KS93} or from Cronin's chiral Lagrangian \cite{C67} as
\be
\lambda \; = \; \bko{K^0}{H_w^{(2)}}{\bar{K}^0} \; = \; 2\,\beta_w^2\,m^2_K \;,
\label{hwkk}
\ee
\begin{figure}[ht]
\epsfxsize=10cm
\centerline{\epsffile{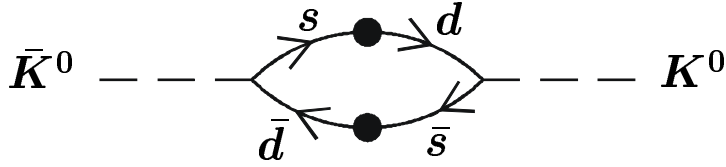}}
\caption{SOW $\bar{K}^0\leftrightarrow K^0$ SQL graph. Each dot represents the
FOW SQL scale $\beta_w$.}
\label{kbark}
\end{figure}
as shown in Fig.~\ref{kbark}. Combining \eqr{hwkk} with \eqr{sintphi}, we find
from data \cite{PDG02,SE95}
\be
\begin{array}{r}
2\,\beta_w^2 \; = \; \frac{\Delta m_{LS}}{m_K} \; = \;
(0.70126\pm0.00121)\times10^{-14} \;\, \\
\Longrightarrow \;\;\; |\beta_w| \; \approx \; 5.921\times10^{-8} \; .
\end{array}
\label{betaw}
\ee
Indeed, this FOW SQL scale $\beta_w$ (determined from the SOW $K_L\!-\!K_S$
mass difference) can be roughly estimated from the GIM $u\,,c$ self-energy
graphs as \cite{GIM70,S81} $|\beta_w|\sim5.6\times10^{-8}$, not far from
\eqr{betaw}.

\begin{figure}[ht]
\epsfxsize=8cm
\centerline{\epsffile{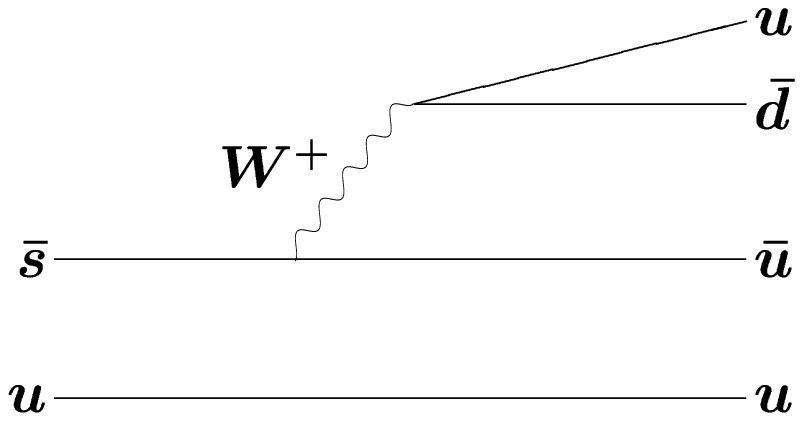}}
\caption{\dit\ W-emission graph.}
\label{we}
\end{figure}

Substituting the FOW scale $\beta_w$ from \eqr{betaw} back into
\eqrs{mkspm}{hwpkl} gives, for $f_K/f_\pi\approx1.22$,
\be
\left|M^{+-}_{K_S}\right|_{\dihs} \; = \; 2\,|\beta_w|\,\frac{f_K}{f^2_\pi} \,
(m^2_K-m^2_\pi) \; \approx \; 35.5\times10^{-8}\;\mbox{GeV} \; .
\label{mkspmd}
\ee
Also, the \dit\ $W$-emission (WE) graph of Fig.~\ref{we} predicts
\be
\!\!\!\!\!\!\left|\bko{\pi^+\pi^0}{H_w}{K^+}\right|_{\mbox{\scriptsize WE}}  = 
\left|\frac{G_F\,V_{ud}\,V_{su}}{2\sqrt{2}}\,(m^2_K\!-\!m^2_\pi)\,f_\pi\right|
\approx  1.86\times10^{-8}\;\mbox{GeV} \; ,
\label{hwppk}
\ee
\noindent near data \cite{PDG02} $|M^{+0}_{K^+}|=(1.83\pm0.01)\times10^{-8}$
GeV. This WE amplitude, when extended to $K_S\!\to\!\pi^+\pi^-$, obviously
doubles \eqr{hwppk} to $2\times1.86\approx3.7\times10^{-8}$ GeV. Then, the
total $K_S\!\to\!\pi^+\pi^-$ amplitude is 
\be
\left|M^{+-}_{K_S}\right|_{\mbox{\scriptsize total}} \; = \;
(35.5+3.7)\times10^{-8} \; \mbox{GeV} \; = \; 39.2\times10^{-8}\;\mbox{GeV} \;,
\label{mkspmt}
\ee
very close to the measured amplitude for $q=206$ MeV \cite{PDG02}, i.e.,
\be
\left|M^{+-}_{K_S}\right|_{\mbox{\scriptsize PDG}} \; = \;
m_K\,\sqrt{\frac{8\pi\Gamma^{+-}_{K_S}}{q}} \; = \; (39.1\pm0.1)\times10^{-8}
\; \mbox{GeV} \; .
\label{mkspmpdg}
\ee

To test this SQL scale (\ref{betaw}) in another way, we note that \eqr{hwpkl}
predicts the FOW scale
\be
\!\!\left|\bko{\pi^0}{H_w}{K_L}\right|  =  \left|\bko{\pi^+}{H_w}{K^+}\right|
 =  2\,\beta_w\,m^2_K\,\frac{f_K}{f_\pi}  \approx  3.58\times10^{-8}
\; \mbox{GeV}^2 \; .
\label{hwpklp}
\ee
The latter scale is compatible with \em eleven \em \/data sets for
$K_S\rightarrow2\pi$, $K\rightarrow3\pi$, $K_L\rightarrow2\gamma$,
$K_L\rightarrow\mu^+\mu^-$, $K^+\rightarrow\pi^+e^+e^-$,
$K^+\rightarrow\pi^+\mu^+\mu^-$, and $\Omega^-\rightarrow\Xi^0\pi^-$, with
data-average value \cite{LS02}
\be
\left|\bko{\pi^0}{H_w}{K_L}\right| \; = \; \left|\bko{\pi^+}{H_w}{K^+}\right|
\; = \; (3.59\pm0.05)\times10^{-8}\;\mbox{GeV}^2 \; .
\label{hpik}
\ee

In passing, we verify the \dih\ rule via our theoretical estimate of the
square of the amplitude in \eqr{mkspmt} relative to the one in \eqr{hwppk}:
\be
\frac{|M^{+-}_{K_S}|\,^2}{|M^{+0}_{K^+}|\,^2} \; = \;
\left(\frac{39.2\ttm{8}\;\mbox{GeV}}{1.86\ttm{8}\;\mbox{GeV}}\right)^2
\; \approx \; 444 \; .
\label{mksdmkp}
\ee
This is near the observed decay-rate ratio 450 found in \eqr{bkskp}.

\section{Indirect and Direct CPV for \bma{K\to2\pi} Decays}
Now we study in detail indirect and direct CPV for $K_{2\pi}$ weak decays.
Given that $|\eta_{+-}/\eta_{00}|$ is close to unity from data \cite{PDG02} in
\eqr{etas}, years ago the leading term $|\varepsilon|$ was thought to be the
$\alpha/\pi$ radiative correction
\be
|\eta_{+-}| \; \approx \; |\eta_{00}| \; \approx \; |\varepsilon| \; \approx \;
\frac{\alpha}{\pi} \; = \; 2.32\times10^{-3} \; ,
\label{alphapi}
\ee
as the origin of CPV \cite{BFL65,CS96}. Actually, $\varepsilon'$
cancels out in \cite{PDG02}
\be
\varepsilon \; = \; \frac{2\eta_{+-}+\eta_{00}}{3} \; = \;
(2.282\pm0.017)\ttm{3} \; ,
\label{epsilon}
\ee
and this is the best measurement of the indirect CPV scale (yet still
compatible with \eqr{alphapi}). Concerning direct CPV, the ratio from
\eqr{etas} gives
\be
\left|\frac{\eta_{+-}}{\eta_{00}}\right| \; \approx \;
\frac{2.286\ttm{3}}{2.274\ttm{3}} \; \approx \; 1.00528 \; ,
\label{etapmzzind}
\ee
a close measure of unity. As a matter of fact, this direct CPV ratio should be
\cite{PDG02}
\be
\left|\frac{\eta_{+-}}{\eta_{00}}\right| \; \approx \;
\left|\frac{\varepsilon+\varepsilon'}{\varepsilon-2\varepsilon'}\right|
\; \approx \; 1+3\,\frac{\varepsilon'}{\varepsilon} \; .
\label{etapmzzdir}
\ee
Equating the CPV data ratio in \eqr{etapmzzind} with the CPV theoretical ratio
in \eqr{etapmzzdir} requires the direct CPV scale
\be 
\Real{\frac{\varepsilon'}{\varepsilon}} \; \approx \; 17.6\ttm{4} \; .
\label{repeth}
\ee
The most recent direct CPV measurements for $K\to\pi\pi$ decays average to
\cite{N02}
\be 
\Real{\frac{\varepsilon'}{\varepsilon}} \; = \; (16.6\pm1.6)\ttm{4} 
\label{repeex}
\ee
(this average includes the new NA48 result, and prior NA31, E731, and KTeV
values). Note that \eqr{repeex} is near the approximate observed CPV scale
in \eqr{repeth}. In fact, the PDG/2002 quotes 
$\Real{\varepsilon'/\varepsilon} \; = \; (18\pm4)\ttm{4}$.

The observed direct CPV is extracted from
\be 
\begin{array}{r}
\displaystyle\Real{\frac{\varepsilon'}{\varepsilon}} \; = \; \frac{1}{6}
\left(\frac{\Gamma_{K_L\to\pi^+\pi^-}\,/\,\Gamma_{K_S\to\pi^+\pi^-}}
           {\Gamma_{K_L\to\pi^0\pi^0}\,/\,\Gamma_{K_S\to\pi^0\pi^0}}
\; - \; 1 \right) \;\, \\[4mm] \displaystyle
\Longrightarrow \;\;\; 6\,\Real{\frac{\varepsilon'}{\varepsilon}} \; = \;
\left(\frac{\eta_{+-}}{\eta_{00}}\right)^2 \; - \; 1 \; .
\end{array}
\label{sixrepe}
\ee
Substituting the latest direct CPV value (\ref{repeex}) into \eqr{sixrepe} in
turn gives
\be
\left|\frac{\eta_{+-}}{\eta_{00}}\right| \; = \; 1.00497\pm0.00048 \; ,
\label{etapmzzex}
\ee
and so a direct CPV scale enhancement of 0.497\% does indeed hint at an
e.m.\ correction to the FOW $K\to2\pi$ amplitudes of Sec.~5,
as we now demonstrate.

\begin{figure}[ht]
\epsfxsize=12cm
\centerline{\epsffile{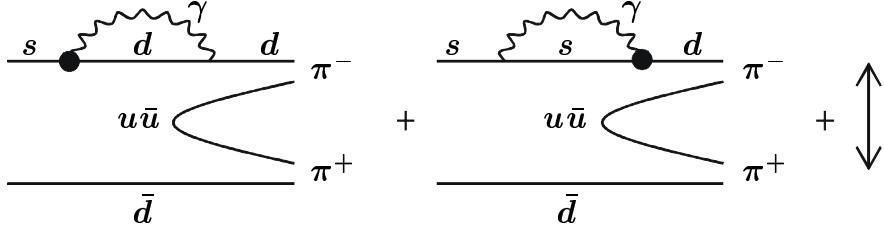}}
\caption{Radiatively corrected \dih\ SQL graphs for $\bar{K}^0\to\pi^+\pi^-$.}
\label{sdgamma}
\end{figure}

In quark-model \dih\ SQL language, there are \em two different \em
\/(see also below) radiatively corrected graphs to be considered for each
charged final-state pion, as depicted in Fig.~\ref{sdgamma}. Then the (direct)
CPV (radiative)
amplitude ratio is predicted to be
\be
\left|\frac{\eta_{+-}}{\eta_{00}}\right|_{\mbox{\scriptsize theory}} \; = \;
1 \, + \, 2\,\frac{\alpha}{\pi} \; \approx \; 1.00465 \; ,
\label{etapmzzth}
\ee
and this 0.465\% enhancement is \em very close \em \/to the 0.497\% direct
CPV enhancement in \eqr{etapmzzex}, and near the 0.528\% approximate
enhancement from \eqr{etapmzzind} in the PDG tables.

First we comment on this crucial factor of 2 in \eqr{etapmzzth}.
As a matter of fact, this was already discussed and explained in \refc{MS83},
having to do with the tadpole transition characterizing the $K\to2\pi$ \dih\
rule. Specifically, the $K^0$ $\bar{s}d$ meson (like the pion) is a tightly
bound (relativistic) Nambu--Goldstone (NG) chiral meson. Then the ``truly
weak'' tadpole ${\mathcal O}(m^2_c/M^2_W)$ \dih\ transition \em cannot \em \/be
transformed away (as can be the loosely bound ``purely e.m.'' transition of
${\mathcal O}(\alpha^2)$ \cite{W73}). Stated another
way, the analog, now measured \cite{A02} $\kappa$(800) meson is generated in a
unitarized coupled-channel approach \cite{BRMDRR86}, but cannot be found in a
loosely bound (nonrelativistic) scheme. This means we must include the factor
of 2 in \eqr{etapmzzth} and Fig.~3, as these two photon-mediated loop graphs
are distinct, since they both occur in NG (tightly bound) kaon configurations.

\begin{figure}[ht]
\epsfxsize=8cm
\centerline{\epsffile{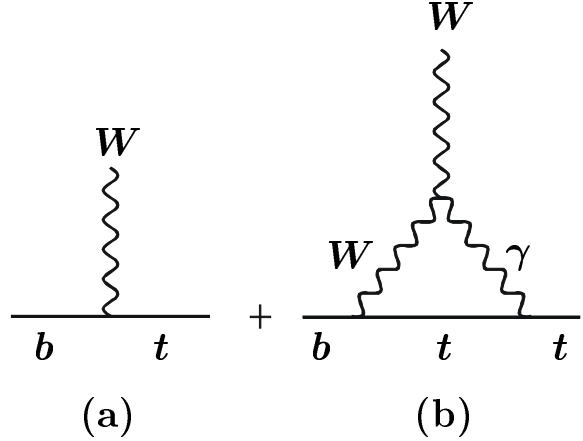}}
\caption{Tree-level (a) and loop-order (b) CKM element $V_{tb}$.}
\label{wwgamma}
\end{figure}

In the context of the SM, the only origin of CPV in the CKM scheme would be in
the $3\times3$ CKM matrix via the phase $\delta$, i.e.\ [note that we use here
the original parametrization due to Kobayashi and Maskawa
(Ref.~\cite{CKM63_73}, second paper; see also the PDG~\cite{PDG02} CKM
review), which is more convenient for our purposes],
\be
V \; = \; \left( \begin{array}{ccc}
V_{ud} & V_{us} & V_{ub}\\ V_{cd} & V_{cs} & V_{cb}\\ V_{td} & V_{ts} & V_{tb}
\end{array}\right) \; = \; \left( \begin{array}{ccc}
c_1 & -s_1 c_3 & -s_1 s_3 \\
s_1 c_2 & c_1 c_2 c_3-s_2 s_3e^{i\delta} & c_1 c_2 s_3+s_2 c_3e^{i\delta}\\
s_1 s_2 & c_1 s_2 c_3+c_2 s_3e^{i\delta} & c_1 s_2 s_3-c_2 c_3e^{i\delta}
\end{array} \right)
\label{ckm}
\ee
which can be written as 
\be
V \; \longrightarrow \; \left( \begin{array}{ccc}
c_1 & -s_1 & 0 \\ s_1 & c_1 & 0 \\ 0 & 0 & -(1+i\delta)
\end{array} \right)
\label{ckmc}
\ee
in the limit of an $SU(4)$ Cabibbo submatrix $\Theta_1=-\Theta_c$, and 
$\Theta_2\,,\Theta_3\to0$. Then \cite{CS96}
\be
-V_{tb} \; = \; 1 + i \lambda_w \frac{\alpha}{\pi}
\ln\left(1+\frac{\Lambda^2}{m_t^2}\right)
\label{vtb}
\ee
follows from the CKM tree and one-loop-order quark graphs of
Fig.~\ref{wwgamma}.
The CPV $WW\gamma\to V\!A\gamma$ vertex is defined as\footnote{
While Marciano \& Queijeiro in \refc{BFL65} assume a parameter $\lambda_w$
(not necessarily unity) in \eqr{vagamma}, we actually assume the weak $V\!-\!A$
CPV e.m.\ coupling $\bk{\gamma_\mu(q)V_\beta}{A_\alpha}=
ie\epsilon_{\alpha\beta\mu\sigma}q^\sigma$ as the natural extension of the 
e.m.\ minimal substitution $q_\mu\to q_\mu\!-\!eA_\mu$, with $\lambda_w$ taken
as unity in an \em a priori \em \/meaning. We let data from \eqr{delta} or
common-sense theory \cite{DGH92} in \eqr{dephi} support our choice. The small
measured bound on an neutron electric dipole moment (NEDM) suggests that it
could play only an insignificant role in the $WW\gamma$ vertex.
}
\be
\bk{\gamma_\mu(q)W_\beta}{W_\alpha} \; = \;
ie\lambda_w\,\epsilon_{\alpha\beta\mu\sigma}\,q^\sigma \; ,
\label{vagamma}
\ee
then generating \eqr{vtb}. The ultraviolet (chiral) cutoff is determined by the
Veltman condition \cite{V81,NSCP89_95}, taking $m_t=174.3$ GeV \cite{PDG02}
$>\!> m_u,m_d,m_s,m_c,m_b$, leading to
\be
\Lambda^2 \; = \; 4m^2_t-(2M^2_W+M^2_Z) \; \approx \; (316.7\;\mbox{GeV})^2\;.
\label{veltman}
\ee
Thus, \eqr{vtb} using the latter chiral cutoff in turn gives from \eqr{veltman}
the phase
\be
\delta \; = \; \lambda_w\frac{\alpha}{\pi}\ln4.301 \; = \;
\lambda_w\,3.389\ttm{3} \; .
\label{deltaw}
\ee
Comparing this CKM version of $\delta$ with the measurement of $\delta$ in
\eqr{delta}, or with the theoretical estimate in \eqr{dephi}, indeed suggests
that the parameter $\lambda_w$ in the $V\!A\gamma$ CPV vertex (\ref{vagamma})
is unity. This parallels the standard e.m.\ minimal substitution
$q_\mu\to q_\mu\!-\!eA_\mu$.

There are, however, various dynamical quark models characterizing the
$V\!A\gamma$ CPV vertex. The Higgs exchange leading to a CPV neutron electric
dipole moment (NEDM) $d_n$ suggested by Weinberg \cite{W76}, giving
$|d_n|\approx4\ttm{24}\:e\!$ cm for $M_H\approx316.7$ GeV, is ruled out by
present data \cite{PDG02} finding $|d_n|<0.6\ttm{25}\:e\!$ cm, as noted by Pal
\& Pham \cite{PP90}. An underlying NEDM version of the $WW\gamma$ CPV coupling
has been often discussed in the literature \cite{nedm}. Now that data appear to
exclude a NEDM origin of CPV, we take the approximate equivalence between
Eqs.~(\ref{delta}), (\ref{dephi}) and the CKM CPV \eqr{deltaw} for
$\lambda_w=1$ as the justification of a $V\!A\gamma$ vertex of unit strength,
in a minimal-substitution sense. Here we should also note that a \em muon \em
\/electric dipole moment (MEDM), now measured as \cite{PDG02} $d_\mu=
(3.7\pm3.4)\ttm{19}\:e\!$ cm, is not ruled out by our theory, which predicts $
d_\mu=(4.2\ttm{21}\:e\,\mbox{cm})\,(m_\mu/2)\,\lambda_w\,\ln(1+\Lambda^2/M^2_W)
\, /\,1$ GeV $=\lambda_w\, 0.62\ttm{21}\:e\!$ cm, for the same chiral cutoff
$\Lambda=316.7$ GeV as in \eqr{veltman}. If we then take $\lambda_w=1$, the
latter MEDM prediction is clearly not excluded by experiment.

So whether we study the quark-model, photon-mediated graphs of Fig.~3
generating the CPV enhancement in $|\eta_{+-}/\eta_{00}|$ of 0.465\%, or the
CKM CPV phase $\delta\approx(3.3\,$--$\,3.4)\ttm{3}$ generated via the
$WW\gamma$
triangle enhancement of Fig.~4, CPV for $K\to2\pi$ decays appears to follow
from photon-mediated loop enhancement. Alternatively, one may compute the small
imaginary corrections to the amplitude for $K^0$-$\bar{K}^0$ transitions,
employing the standard double-$W$ box diagrams, but accounting for the
physical decay thresholds of the $W$ and sandwiching the whole amplitude
between composite-kaon vertex functions, which results in an effect
surprisingly close to the value of $|\varepsilon|$ \cite{BR03}. However, the
latter $WW$-box result is the same as the indirect CPV value in the present SQL
$K^0$-$\bar{K^0}$ scheme of Fig.~1 and Eqs.~(\ref{hwkk},\ref{betaw}).\footnote{
The SQL SOW $K^0$-$\bar{K}^0$ scheme in Fig.~1 correctly treats the chiral
$K^0$=$\bar{s}d$ state as a tightly bound (Nambu--Goldstone) meson (whence the
crucial factor of 2 for CPV in \eqr{etapmzzth}). Although the alternative
$WW$-box SOW $K^0$-$\bar{K}^0$ approach in principle involves many additional
graphs, in the CL the latter formulation should reduce to the former SQL
scheme, leading to the $K_S\to2\pi$ amplitudes in
Eqs.~(\ref{mkspmd},\ref{hwppk},\ref{mkspmt}).
}

Another weak-interaction radiative correction of this small size follows from
the semileptonic weak $\mu^-\to e^-\nu_\mu\bar{\nu}_e$ decay. Long ago,
Kinoshita \& Sirlin \cite{KS59_78} computed this net $V\!-\!A$ rate correction 
due to radiative effects as
\be
\frac{\Gamma-\Gamma_0}{\Gamma_0} \; = \; \frac{\alpha}{2\pi}\,\left(\pi^2-
\frac{25}{4}\right) \; \approx \; 0.42\,\% \; .
\label{vma}
\ee
We note that this 0.42\% radiative enhancement is near the 0.465\% e.m.\
enhancement for $K\to2\pi$, or the observed direct CPV scale of 0.497\%.

\section{Summary and Conclusions}
In the foregoing, we have analyzed remarkable patterns in the experimental data
for CPC and CPV \ktp\ and \ktpg\ weak decays, as well as for \dih\ compared to
\dit\ processes. Concretely, 
in Sec.~2 we extracted from data \cite{PDG02} the 4.5\% to 4.7\% \dit\
suppression in both CPC and CPV $K\to2\pi$ transitions. Next, in Sec.~3 we
found that the observed $K\to\pi\pi\gamma/\pi\pi$ branching ratios are all
scaled near $\alpha/\pi$ or $\alpha/2\pi$. Then in Sec.~4 we studied the CPV
angles $\phi$ and $\delta$. In Sec.~5 we returned to $K\to2\pi$ \dih\
dominance from the perspective of the chiral (constituent) quark model.
Finally, in Sec.~6 we studied indirect and direct kaon CPV in the framework of
photon-mediated loop graphs via 0.497\%, 0.465\%, 0.42\% values, obtained from
direct CPV data, and theoretical radiatively corrected \ktp\ and
$\mu\to e\nu\bar{\nu}$ weak amplitudes.

We have not considered strong-interaction penguin graphs in our analysis, as
the typical QCD scale of $\sim$ 1 fm is orders of magnitude larger than the
electroweak CPV scale of about $0.0023\times1/M_W=5.6\times10^{-6}$ fm.
Therefore,
we argue it is unlikely that such graphs yield a significant contribution to
CPV-related processes. This qualitative argument is confirmed by explicit
calculations, showing that QCD penguins lead to much too small results for
\dih\ nonleptonic kaon decay rates \cite{CFGHR86_80}, as well as  for the
CPV ratio $\epsilon'/\epsilon$ \cite{BJL93_95}. Also Lucio \cite{L81} showed,
using inner-bremsstrahlung and direct-emission graphs, that QCD penguins do
not play an important role in \dih\ $\,K\to\pi\pi\gamma$ transitions.

Summarizing, we have shown that CPV --- at least for kaon weak decays
involving two pions --- can be described with standard second-order weak
physics, radiatively corrected. 
We believe this line of research should be further pursued for other CPV
processes, too, as a possible alternative to new physics.

To conclude, we present CPV to CPC decay-rate ratios for various PDG
\cite{PDG02} kaon-decay processes involving two pions, as extra support
for our analysis: \\[-5mm]
\bea
\frac{\Gamma_{K_L\,\to\,\pi^+\pi^-}}{\Gamma_{K_S\,\to\,\pi^+\pi^-}}
\;\; & = & \;\; (5.25\pm0.12)\times10^{-6} \; , \label{ktwopi1} \\
\mbox{close to} \;\;\;|\eta_{+-}|^2 \;\; & = & \;\; (5.23\pm0.08)\times10^{-6}
\; ; \label{ktwopi2} \\[2mm]
\frac{\Gamma_{K_L\,\to\,\pi^0\pi^0}}{\Gamma_{K_S\,\to\,\pi^0\pi^0}}
\;\; & = & \;\; (5.19\pm0.20)\times10^{-6} \; , \label{ktwopi3} \\
\mbox{close to} \;\;\;|\eta_{00}|^2 \;\; & = & \;\; (5.17\pm0.08)\times10^{-6}
\; ; \label{ktwopi4} \\[2mm]
\frac{\Gamma_{K_L\,\to\,\pi^+\pi^-e^+e^-}}{\Gamma_{K_S\,\to\,\pi^+\pi^-e^+e^-}}
\;\; & = & \;\; (13.4\pm3.3)\times10^{-6} \; ; \label{ktwopi5} \\[2mm]
\frac{\left(\Gamma_{K_L\,\to\,\pi^+\pi^-\gamma}\right)_{\mbox{\scriptsize CPV}}}
{\Gamma_{K_S\,\to\,\pi^+\pi^-\gamma}} \;\; & = & \;\; |\eta_{+-\gamma}|^2
\;\; = \;\; (5.52\pm0.33)\times10^{-6} \; . \label{ktwopi6} 
\eea
Note that the experimental rate ratio in \eqr{ktwopi5} has a large error, 
suggesting that it might also turn out to be close to the average value of
$(5.2\pm0.1)\times10^{-6}$ in Eqs.~(\ref{ktwopi1})--(\ref{ktwopi6}), being
about 2.5 standard deviations away. Finally, as the CPV component of the
process $K_L\to\pi^+\pi^-\gamma$ is not yet known experimentally, we have used
in \eqr{ktwopi6} the equality of the corresponding rate ratio with the
measured quantity $|\eta_{+-\gamma}|^2$. Clearly, more and still better
experiments, also in the kaon sector, are needed to lend additional evidence to
our interpretation of CPV as a radiatively corrected SOW effect.  \\[3mm]

\noindent{\large\bf Acknowledgments} \\[1mm]
We wish to thank F.~Kleefeld for valuable discussions.
This work was partly supported by the
{\it Funda\c{c}\~{a}o para a Ci\^{e}ncia e a Tecnologia}
of the {\it Minist\'{e}rio da Ci\^{e}ncia e do Ensino Superior} \/of Portugal,
under contract no.\ POCTI/\-FNU/\-49555/\-2002.

\end{document}